\documentclass[12pt]{article}
\usepackage[table,xcdraw]{xcolor}
\usepackage{amsmath}
\usepackage{adjustbox}
\DeclareMathOperator{\sech}{sech}
\DeclareMathOperator{\csch}{csch}

\DeclareMathOperator{\arcsinh}{arcsinh}

\usepackage{amsmath,bm}
\usepackage{url}
\usepackage{tikz}
\usetikzlibrary{shapes.geometric}
\usetikzlibrary{arrows}
\usepackage{amssymb}
\usepackage{rotating}
\usepackage{physics}
\usepackage{epsfig}
\usepackage{graphicx}
\usepackage{color}
\usepackage{cite}
\usepackage{subfig}
\usepackage{slashed}
\usepackage{mathtools}
\usepackage{lipsum}
\usepackage[linktocpage]{hyperref}
\usepackage{hhline}
\usepackage[font=small,labelfont=bf]{caption}

\newcommand{\beq}{\begin{equation}}
\newcommand{\eeq}{\end{equation}}
\newcommand{\ga}{\lower.7ex\hbox{$\;\stackrel{\textstyle>}{\sim}\;$}}
\newcommand{\la}{\lower.7ex\hbox{$\;\stackrel{\textstyle<}{\sim}\;$}}

\hypersetup{
    colorlinks = true,
    citecolor = {blue},
    linkcolor = {blue},
    urlcolor = {blue},
}

\begin{document}

\def\jcap{\ref@jnl{J. Cosmology Astropart. Phys.}}


\begin{center}
{\bf {\large Quintessential Inflation in Palatini Gravity}}

\end{center}

\begin{center}{
{\bf Sarunas~Verner}}
\end{center}

\begin{center}
{\em William I. Fine Theoretical Physics Institute, School of
 Physics and Astronomy, University of Minnesota, Minneapolis, MN 55455,
 USA}
 
 \end{center}

\vspace{0.3cm}
\centerline{\bf ABSTRACT}
\vspace{0.1cm}

    {\small We study a model of quintessential inflation in the context of Palatini gravity. As a representative example, we consider the Peebles-Vilenkin model of quintessential inflation with a small non-minimal coupling to gravity, which is consistent with the most recent Planck measurements. At the end of inflation, the inflaton field passes through a tachyonic region and it leads to explosive particle production through the tachyonic preheating process. After preheating, the Universe becomes dominated by the kinetic energy of the inflaton and enters a period of kination. Eventually, the total energy density of the Universe becomes dominated by radiation, resulting in reheating. We find that the model predicts the reheating temperature values $T_{\rm{RH}} \sim \mathcal{O}(10^3 - 10^8) \, \rm{GeV}$, which is significantly above the temperature of Big Bang Nucleosynthesis.  Following reheating, the inflaton field rolls down the quintessence potential until it freezes. Since the quintessence remains frozen until the present day, the residual potential energy density at this field value explains the observed dark energy density.
} 

\vspace{0.2in}

\begin{flushleft}
{October} 2020
\end{flushleft}
\medskip
\noindent

\newpage
\noindent\rule{\textwidth}{0.4pt}
\vspace{-1.2cm}
{
  \tableofcontents
}
\noindent\rule{\textwidth}{0.4pt}

\section{Introduction}
An early epoch of accelerated expansion of the Universe called cosmic inflation is the leading theory explaining the near-flatness and homogeneity of the Universe~\cite{Olive:1989nu, Linde:2005ht, Lyth:1998xn, Martin:2013tda, Martin:2013nzq, Martin:2015dha}. Although the complete picture of inflationary paradigm is not yet fully understood, the usual scenarios assume that the exponential expansion of the Universe is driven by a single (or multiple) scalar fields in curved spacetime. Following the inflationary epoch, the scalar field starts oscillating about the minimum of its effective potential and ultimately decays into elementary particles and reheats the Universe~\cite{Dolgov:1982th, Abbott:1982hn, Nanopoulos:1983up}.

As is well known, the slowly-rolling inflaton field leads to
the scalar field fluctuations that are predominantly Gaussian, with a tilted scalar perturbation spectrum, $n_s \neq 1$, and a small ratio between the primordial gravitational wave and the primordial curvature perturbation power spectra, $r \ll 1$. Present measurements of the cosmic microwave background (CMB) radiation from the Planck satellite determine the amplitude of scalar perturbations to be $A_s \simeq 2.1 \times 10^{-9}$ and that the scalar spectrum shows a presence of small non-Gaussianity, with a spectral tilt $n_s \simeq 0.965$~\cite{Aghanim:2018eyx,Akrami:2018odb}, which strongly supports the inflationary paradigm. The BICEP2/Keck Array measurements also place stringent upper bound on the tensor-to-scalar ratio, $r < 0.06$~\cite{Ade:2018gkx}. Future CMB experiments such as LiteBIRD~\cite{Matsumura:2013aja}, SO~\cite{Ade:2018sbj}, PICO~\cite{Hanany:2019lle}, and CMB-S4~\cite{Abazajian:2016yjj} are projected to measure the limit on tensor-to-scalar ratio to an accuracy of $r \simeq 0.001$, allowing to discriminate between different models of inflation.

The dynamics of inflation is usually studied in the \textit{metric} formulation of general relativity, where it is assumed that the space-time connection $\Gamma$ is metric-compatible, which in turn implies that it can be uniquely determined by the metric tensor only. However, in the \textit{Palatini} theory of gravity (see~\cite{Tenkanen:2020dge} for a recent review), the space-time connection $\Gamma$ and the metric 
tensor $g_{\mu \nu}$ are considered to be independent variables. It was first pointed out by Bauer and Demir~\cite{Bauer:2008zj} (see also~\cite{Bauer:2010jg, Tamanini:2010uq}) that non-minimally coupled scalar field theories have different inflationary dynamics in metric and Palatini theories of gravity, mainly because the models have a different dependence on the non-minimal coupling arising from conformal transformations, and in turn they lead to different predictions of the cosmological observables $n_s$ and $r$.


The inflationary models that have been excluded by the Planck data become viable again in theories with a non-minimal coupling to gravity, including models based on simple monomial scalar potential~\cite{Tenkanen:2017jih, Antoniadis:2018yfq}.
Recently, cosmic inflation in Palatini formulation has been studied in different contexts~\cite{Fu:2017iqg, Jarv:2017azx, Racioppi:2018zoy, Carrilho:2018ffi, Enckell:2018hmo, Antoniadis:2018ywb, Almeida:2018oid, Takahashi:2018brt, Jinno:2018jei, Edery:2019txq, Giovannini:2019mgk, Tenkanen:2019wsd, Bostan:2019wsd, Gialamas:2019nly, Racioppi:2019jsp, Tenkanen:2020cvw, Lloyd-Stubbs:2020pvx, Antoniadis:2020dfq, Ghilencea:2020piz, Takahashi:2020car, Das:2020kff, Jarv:2020qqm, Gialamas:2020snr, Karam:2020rpa}, including appealing minimal scenarios where the Standard Model Higgs boson is associated with the inflaton field~\cite{Bezrukov:2007ep},\footnote{For a recent review on Higgs inflation, see~\cite{Rubio:2018ogq}} which are known as Palatini Higgs inflation models~\cite{Bauer:2008zj, Bauer:2010jg, Rasanen:2017ivk, Racioppi:2017spw, Markkanen:2017tun, Enckell:2018kkc, Kannike:2018zwn, Rasanen:2018fom, Rasanen:2018ihz, Tenkanen:2019jiq, Rubio:2019ypq, Jinno:2019und, Tenkanen:2019xzn, Shaposhnikov:2020geh, Shaposhnikov:2020fdv, McDonald:2020lpz, Gialamas:2020vto}. Usually, models of inflation in Palatini formulation have significantly smaller tensor-to-scalar ratio, $r$, and the upcoming CMB B-mode experiments may be used to distinguish between metric and Palatini formulations in non-minimally coupled models.

In this paper, we study the original model of quintessential inflation, first proposed by Peebles and Vilenkin~\cite{Peebles:1998qn}, in the context of Palatini gravity. The inflationary epoch of this model is approximated by a quartic scalar potential, which is perfectly consistent with current Planck measurements in theories with a non-minimal coupling to gravity, as discussed in the previous paragraph. After the end of inflationary epoch, the scalar inflaton field rolls down its potential until it freezes at the field value, $\phi_{F} \sim M_P$. Since the field remains frozen until the present day, the effective potential energy density $U(\phi_F)$
acts as dark energy and explains the current accelerating expansion of the Universe.

Such a model of quintessential inflation is a typical model of non-oscillatory inflation, where at the end of inflation the inflaton field does not oscillate about its scalar potential minimum. Usually, these models are plagued by a very ineffective reheating mechanism through the gravitational particle production~\cite{Parker:1968mv, Grib:1976pw, Zeldovich:1977vgo, Ford:1986sy, Hashiba:2018iff}, which generally leads to very low reheating temperatures. However, the situation can be remedied if the instant preheating mechanism~\cite{Felder:1999pv, Felder:1998vq} is considered instead.

The non-minimal models of inflation in metric and Palatini formulations reheat the Universe through different mechanisms. In the metric case, the Universe is usually reheated through the $\textit{parametric resonance}$~\cite{Kofman:1994rk, Shtanov:1994ce, Kofman:1997yn, Greene:1997fu}, whereas in the Palatini case the particle production is dominated by the \textit{tachyonic preheating}~\cite{Felder:2000hj, Felder:2001kt}, and both models were studied recently in the context of Palatini Higgs inflation~\cite{Rubio:2019ypq, Karam:2020rpa}.
Because the tachyonic preheating leads to an extremely efficient exponential growth of scalar field fluctuations, it is a suitable reheating mechanism for models of quintessential inflation.

The structure of this paper is as follows. In Section~\ref{sec:formulation} we review general features of Palatini formulation of gravity. Then, in Section~\ref{sec:slowroll} we introduce the slow-roll approximation and compare the cosmological parameters between the metric and Palatini formulation. Section~\ref{sec:quintessence} introduces a model of quintessential inflation in the context of Palatini gravity. A detailed analysis of the preheating process for our model is presented in Section~\ref{sec:preheating}.
We calculate the reheating temperature and estimate the abundance of gravitational waves in Section~\ref{sec:reh}. Finally, we summarize our results and conclude in Section~\ref{sec:results}. 

\section{Palatini Formulation}
\label{sec:formulation}
The most commonly used framework to study the inflationary epoch of the Universe is based on the \textit{metric} formulation of general relativity. In metric formalism, one assumes that the space-time connection is torsion-free,{\footnote{For models that do not assume vanishing torsion, see~\cite{Rasanen:2018ihz, Shimada:2018lnm}.}} $\Gamma_{\mu \nu}^{\rho} =  \Gamma_{\nu \mu}^{\rho}$, and metric-compatible, $\nabla_{\rho} \, g_{\mu \nu} = 0$, where $\nabla_{\rho}$ is a covariant derivative. These two postulates
determine that the connection is characterized only by the metric tensor, $g_{\mu \nu}$.

However, in the \textit{Palatini} formulation of general relativity,\footnote{For a recent review on Palatini gravity, see~\cite{Tenkanen:2020dge}.} both the space-time metric $g_{\mu \nu}$ and the connection $\Gamma$ are treated as independent variables, and the metric compatibility is no longer assumed, in contrast to the metric theory. For a more detailed discussion of two different formulations of gravity, see~\cite{Bauer:2008zj} and references therein. Importantly, in theories with \textit{non-minimal} coupling to gravity, the metric and Palatini approaches lead to different predictions of the principal CMB observables~\cite{Aghanim:2018eyx, Akrami:2018odb, Ade:2018gkx}, such as the scalar spectral tilt, $n_s$, and the tensor-to-scalar ratio, $r$. The inflationary models based on the Palatini theory of gravity predict the value of $r$ to be inversely proportional to the coupling constant, $\xi$, (see Eq.~(\ref{eq:palacons1})), and for $\xi \gg 1$, the value of $r$ is many orders of magnitude smaller than its metric counterpart, and we discuss it in detail in the next Section. 

In the Lagrangian formulation of general relativity, the Jordan frame action is given by:
\begin{equation}
    \label{eq:actionjord}
    \mathcal{S}_{J} \; = \; \int d^4x \sqrt{-g} \left[\frac{1}{2} M_P^2 \, \Omega^2(\phi) g^{\mu \nu} R_{\mu \nu}(\Gamma)  - \frac{1}{2} K_J(\phi) \nabla_{\mu} \phi \nabla^{\mu} \phi - V (\phi) \right],
\end{equation}
where $M_P = 2.4 \times 10^{18} \, \rm{GeV}$ is the reduced Planck mass, $\Omega(\phi)$ is the conformal factor, $K_J(\phi)$ is the kinetic term, $V(\phi)$ is the effective scalar potential, and the scalar curvature is defined as:
\begin{equation}
    \label{eq:scalarcurv}
    R \; \equiv \; g^{\mu \nu} R_{\mu \nu} (\Gamma , \, \partial \Gamma) \, ,
\end{equation}
which implicitly assumes that the connection is metric-compatible in \textit{metric} theory, or that $g_{\mu \nu}$ and $\Gamma$ are independent variables in $\textit{Palatini}$ theory.\footnote{It is still assumed that in Palatini theory the connection is torsion-free.} For the remainder of this paper, we set $K_J = 1$.

Since the Jordan frame action~(\ref{eq:actionjord}) contains a \textit{non-minimal} coupling to gravity,
it is conventional to transform this action and study the models of inflation in the canonical Einstein frame, where $\Omega^2(\phi) = 1$. The transformation between the Jordan and Einstein frames can be achieved by performing the following conformal transformation:
\begin{equation}
    \label{eq:conftrans1}
    \tilde{g}_{\mu \nu} (\phi) \equiv \Omega^2(\phi) g_{\mu \nu} (\phi),
\end{equation}
which leads to
\begin{equation}
    \label{eq:conftrans2}
    \sqrt{-g} \; = \; \Omega^{-4} \sqrt{-\tilde{g}},
\end{equation}
and 
\begin{equation}
    \label{eq:conftrans3}
    R  \; = \; \Omega^2 \left[\tilde{R} + c \times \left( \frac{6 \widetilde{\Box}{\Omega}}{\Omega} -  \frac{12 \widetilde{\nabla}_{\mu} \Omega \widetilde{\nabla}^{\mu} \Omega }{\Omega^2} \right) \right] \, ,
\end{equation}
where $c = 1$ for the \textit{metric} formulation and $c = 0$ for the \textit{Palatini} formulation, and the quantities with a tilde are given in terms of the metric $\tilde{g}_{\mu \nu}$. If we apply the conformal transformations~(\ref{eq:conftrans1})-(\ref{eq:conftrans3}) to the action~(\ref{eq:actionjord}) and use the integration by parts, we find the Einstein frame action:
\begin{equation}
    \label{eq:actioneinst}
    \mathcal{S}_E \; = \; \int d^4 x \sqrt{- \tilde{g}} \left[ \frac{1}{2} M_P^2 \tilde{R}^2 - \frac{1}{2} K_E(\phi) \widetilde{\nabla}_{\mu} \phi \widetilde{\nabla}^{\mu} \phi - \frac{V(\phi)}{\Omega^4}
      \right] \, ,
\end{equation}
where
\begin{equation}
    \label{eq:kinetic}
    K_E \; = \; \frac{1}{\Omega^2} + c \times \frac{6 \, \Omega_{\phi}^2 }{\Omega^2} \, M_P^2,
\end{equation}
with $\Omega_{\phi} \equiv \partial{\Omega}/\partial{\phi}$. 

To express the kinetic term~(\ref{eq:kinetic}) in a canonical form, we introduce the field redefinition
\begin{equation}
    \label{eq:canonical}
    \frac{d \phi}{d \chi} = \sqrt{\frac{\Omega^2}{1 + c \times 6 \Omega_{\phi}^2 M_P^2}} \, ,
\end{equation}
and the Einstein frame action~(\ref{eq:actioneinst}) becomes
\begin{equation}
    \mathcal{S}_E \; = \; \int d^4 x \sqrt{-\tilde{g}} \left[\frac{1}{2} M_P^2 \tilde{R}^2 - \frac{1}{2} \widetilde{\nabla}_{\mu} \chi \widetilde{\nabla}^{\mu} \chi - U(\chi) \right] \, ,
\end{equation}
where
\begin{equation}
    \label{eq:potcan}
    U(\chi) \equiv \frac{V(\phi(\chi))}{\Omega^4(\phi(\chi))},
\end{equation}
is the canonically-normalized Einstein frame scalar potential.

Note that the Palatini and metric formulations have different
canonical field redefinition forms~(\ref{eq:canonical}), which affects the effective scalar potential in the Einstein frame~(\ref{eq:potcan}), and in turn changes the inflationary dynamics.

\section{Slow-Roll Inflation}
\label{sec:slowroll}

As was discussed in the previous Section, the main difference between the metric and Palatini formulations of gravity arises from the conformal transformation of the scalar curvature~(\ref{eq:conftrans3}), where in the Palatini case $R \rightarrow \Omega^2(\phi) R$.

In this paper, we study the Peebles-Vilenkin model of \textit{quintessential inflation} that is based on the quartic inflationary potential~\cite{Peebles:1998qn}, which is excluded by the Planck measurements in theories with a \textit{minimal} coupling to gravity. However, this model is perfectly compatible with the observed data if instead we consider a \textit{non-minimal} coupling to gravity. Here we explore the inflationary dynamics of the quartic potential\footnote{One may also consider other inflationary models based on monomial potentials with the inflaton field coupled \textit{non-minimally} to the scalar curvature, e.g., see~\cite{Tenkanen:2017jih, Bostan:2019wsd} for the quadratic inflation model with
a non-minimal coupling to gravity.} and discuss the evolution from inflation to the present day in the Sections below.

We consider the following quartic inflationary potential:
\begin{equation}
    \label{eq:quartic}
    V(\phi) = \frac{\lambda}{4} \phi^4 \, ,
\end{equation}
and we introduce the conformal factor
\begin{equation}
    \label{eq:conffact1}
    \Omega^2(\phi) = 1 +  \frac{\xi \phi^2}{M_P^2} \, ,
\end{equation}
where $\xi$ is a dimensionless coupling constant.\footnote{See~\cite{Linde:2011nh, Kallosh:2013hoa, Kallosh:2013pby} for the studies of chaotic inflation with a non-minimal coupling to gravity.} For the remainder of this paper we set $M_P = 1$ and omit the factor of $M_P$ in most analytical expressions. For the models of inflation in Palatini formalism that consider different conformal factors, see, e.g.,~\cite{Jarv:2017azx, Carrilho:2018ffi, Enckell:2018hmo, Antoniadis:2018ywb, Jinno:2019und}. Combining the conformal factor~(\ref{eq:conffact1}) with Eq.~(\ref{eq:canonical}), we find

\begin{equation}
    \label{eq:kinetic2}
    \frac{d \phi}{d \chi} \; = \; \frac{1 + \xi \phi^2}{\sqrt{1 + \xi  \phi^2 \, (1 + 6 \, c \, \xi)}},
\end{equation}
and if we integrate this expression to find the canonical field redefinition, we obtain~\cite{Bauer:2008zj, GarciaBellido:2008ab, Rasanen:2017ivk}:
\begin{equation}
    \label{eq:kineticcanon1}
     \phi(\chi) \simeq \frac{1}{\sqrt{\xi}} \exp(\sqrt{\frac{1}{6}} \chi), \qquad (\rm{Metric}) \, ,
\end{equation}
\begin{equation}
    \label{eq:kineticcanon2}
    ~~\phi(\chi) = \frac{1}{\sqrt{\xi}} \sinh(\sqrt{\xi} \chi), \qquad ~(\rm{Palatini}) \, ,
\end{equation}
where in the metric case we assume that $\xi \gg 1$, and the expression is exact in the Palatini formulation. Using the field redefinitions~(\ref{eq:kineticcanon1})-(\ref{eq:kineticcanon2}), we find that the Einstein-frame potential~(\ref{eq:potcan}) in terms of the canonically-normalized field, $\chi$, can be expressed as
\begin{align}
\label{potmetric}
U(\chi) & \simeq  \frac{\lambda}{4 \xi^{2}}\left(1+\exp \left(-\sqrt{\frac{2}{3}} \, \chi \right)\right)^{-2}, \,  ~~\text{(Metric)} \, , \\ 
\label{potpalatini}
U(\chi) &=  \frac{\lambda}{4 \xi^{2}} \tanh ^{4}\left(\sqrt{\xi} \chi \right), \qquad \qquad \qquad \text{(Palatini)} \, ,
\end{align}
where both forms describe the plateau-like potentials that are suitable for \textit{slow-roll} inflationary dynamics. To characterize the period of inflation with a slowly-rolling field, we neglect the acceleration term $\ddot{\chi}$ and approximate the Klein-Gordon equation as:
\begin{equation}
    \label{eq:KG}
    3 H \dot{\chi} + U'(\chi) \simeq 0 \, ,
\end{equation}
where $U'(\chi) = \partial U(\chi)/\partial \chi$ and $H \equiv \dot{a}/a$ is the Hubble expansion rate. From the Friedmann equation, 
$H^2 = \rho_{\chi}/3$, we find that the Hubble expansion rate is dominated by its potential energy contribution,
\begin{equation}
    \label{eq:hubblefried}
    3H^2 \simeq U(\chi).
\end{equation}
To describe the dynamics of inflation driven by a single scalar field, we introduce the slow-roll parameters $\epsilon$ and $\eta$:
\begin{equation}
    \label{eq:slowrollpar}
    \epsilon \equiv \frac{1}{2} \left(\frac{U^{\prime}(\chi)}{U(\chi)}\right)^{2} , \qquad \eta \equiv \frac{U^{\prime \prime}(\chi)}{U(\chi)},
\end{equation}
where in the slow-roll approximation we must satisfy the conditions $\epsilon \ll 1$ and  $|\eta| \ll 1$,
requiring the inflationary potential to be flat. It is important to note that the second derivative of the potential $U''(\chi)$ can be identified with the effective inflaton mass, which must be significantly smaller than the Hubble rate that is constant during inflation. For 
the discussion beyond the slow-roll approximation, see~\cite{Lyth:1998xn}.

The slow-roll approximation can be used to 
determine the Hubble expansion rate in terms of the scalar field value. The expansion rate can be expressed in terms of the number of \textit{e-folds}:
\begin{equation}
N_{*} \equiv \ln \left(\frac{a_{\mathrm{end}}}{a_{*}}\right)=\int_{t_{*}}^{t_{\mathrm{end}}} H d t \simeq-\int_{\chi_{*}}^{\chi_{\mathrm{end}}} \frac{1}{\sqrt{2 \epsilon}} d \chi,
\end{equation}
where the inflaton field value at the end of inflation, $\chi_{\rm{end}}$, is defined by $\epsilon(\chi_{\rm{end}}) = 1$ and $\chi_{*}$ is the field value 
when the pivot scale $k_*$ exits the horizon. The number of e-folds between the horizon exit scale and the end of inflation is given by $N_*$. To solve the flatness and horizon problems, typical values of $N_*$ lie in the range $\sim 50 - 60$~\cite{Liddle:2003as, Martin:2010kz}.

To make the connection between models of inflation and CMB observables~\cite{Aghanim:2018eyx, Akrami:2018odb, Ade:2018gkx}, we express the scalar spectral tilt, $n_s$, and the tensor-to-scalar ratio, $r$, in terms of the slow-roll parameters:
\begin{eqnarray}
{\rm Amplitude~of~scalar~perturbations}~A_s:\; 
A_s \;& = &\; \frac{U(\chi_*)}{24 \pi^2 \epsilon_*} \simeq 2.1 \times 10^{-9} \, , \label{eq:As} \\ \notag
\hspace{-15mm} {\rm Scalar~spectral~tilt}~n_s:\;  n_s \; & \simeq &\; 1 - 6 \epsilon_* + 2 \eta_*\\
& = &\; 0.965 \pm 0.004 \; (68\%~{\rm CL}) \, ,
\label{eq:ns} \\
\hspace{-5mm} {\rm Tensor\mbox{-}to\mbox{-}scalar~ratio}~r:\;  r \; &
\simeq & \; 16 \, \epsilon_* < 0.061 \; \,  (95\%~{\rm CL}) \, , \label{eq:r}
\label{observables}
\end{eqnarray}
where these parameters are evaluated at the pivot scale $k_* = 0.05 \, \rm{Mpc}^{-1}$.

One can now readily calculate the predictions for the amplitude of scalar perturbations, $A_s$, the tilt in the spectrum of scalar perturbations, $n_s$, and the tensor-to-scalar ratio, $r$, for different metric and Palatini Einstein frame potentials~(\ref{potmetric})-(\ref{potpalatini}):
\begin{equation}
\label{cmbmetric}
n_{s} \simeq 1-\frac{2}{N_{*}}, \qquad 
r \simeq \frac{12}{N_*^2}, \qquad A_s \simeq \frac{\lambda N_{*}^{2}}{72 \pi^{2} \xi^{2}}, \quad (\text{Metric}) \, ,
\end{equation}
\begin{equation}
\label{cmbpalatini}
n_{s} \simeq 1-\frac{2}{N_{*}}, \qquad r \simeq \frac{2}{\xi N_{*}^{2}}, \qquad A_s \simeq \frac{\lambda N_{*}^{2}}{12 \pi^{2} \xi}, \quad (\text{Palatini}) \, ,
\end{equation}
where $\xi \gg 1$. The most important difference between the metric and Palatini theories is the predicted tensor-to scalar ratio, where in the latter case, the values of $r$ are inversely proportional to the coupling constant, $\xi$. Present CMB measurements set upper limits on the tensor-to-scalar ratio, $r \lesssim 0.06$~\cite{Ade:2018gkx}, which in Palatini formalism corresponds to a coupling constant $\xi \gtrsim 33.3/N_*^2$. 

If we use the observed value of the amplitude of scalar perturbations, $A_s \simeq 2.1 \times 10^{-9}$,
with Eqs.~(\ref{cmbmetric})-(\ref{cmbpalatini}) we find the relations between the non-minimal coupling constant $\xi$ and inflaton coupling $\lambda$:
\begin{equation}
    \label{eq:metcons1}
    ~~\frac{\xi}{\sqrt{\lambda}} \simeq 800 \, N_*, \qquad (\text{Metric}) \, ,
\end{equation}
\begin{equation}
    \label{eq:palacons1}
    \frac{\xi}{\lambda} \simeq 4 \times 10^6 \, N_*^2, \qquad (\text{Palatini}) \, .
\end{equation}
Assuming that $\lambda < 0.1$, we find the constraints
\begin{equation}
    \label{eq:cons}
    \xi \lesssim 250 \, N_*, \qquad (\text{Metric}) \, , \qquad \xi \lesssim 4 \times 10^5 \, N_*^2 \, , \qquad (\text{Palatini}) \, ,
\end{equation}
and for $N_* \simeq 50 - 60$ we obtain $\xi_{\rm{max}} \simeq (1.3 - 1.5) \times 10^{4}$ in the metric case and $\xi_{\rm{max}} \simeq (1.0 - 1.4) \times 10^{9}$ in the Palatini case.

Future CMB experiments, such as LiteBIRD~\cite{Matsumura:2013aja}, SO~\cite{Ade:2018sbj}, PICO~\cite{Hanany:2019lle}, and CMB-S4~\cite{Abazajian:2016yjj} will constraint the tensor-to-scalar ratio with an upper bound $r \lesssim 0.001$. 
If the primordial gravitational waves are not observed above this limit, then the metric models of inflation with a non-minimal coupling to gravity, that predict $r \sim 0.003 - 0.005$, would be excluded. Therefore, the limit $r \sim 0.001$ serves as an important target that may be used to distinguish between different formulations of gravity.

From Eq.~(\ref{eq:cons}) we see that the values of $r$ in the Palatini formalism can be as low as $r \sim \mathcal{O}(10^{-13})$, and in the absence of gravitational wave detection, Palatini models of inflation would still be consistent with the observed data. Importantly, 
models of quintessential inflation with a non-minimal coupling to gravity have a significantly smaller maximum coupling constant, with $\xi_{\rm{max}} \sim \mathcal{O}(1)$,
corresponding to the tensor-to-scalar ratio $r \sim \mathcal{O}(10^{-4})$, and these models could be confirmed or ruled out by next-generation CMB experiments. We discuss it in detail in subsequent Sections.

It should be noted that the expressions~(\ref{cmbpalatini}) are only suitable for $\xi \gg 1$. However, when we consider models of quintessential inflation in Palatini gravity with $\xi \lesssim \mathcal{O} (1)$, we use the full expressions of CMB observables, given by
\begin{equation}
    \label{cmbpalatinifull}
    n_s = \,   \frac{1 + \kappa (N - 1) - 8\xi N(N-2)}{N(\kappa - 8\xi N ) -1}, \, \qquad r = \frac{16}{1 - \kappa N + 8 \xi N^2} \qquad (\text{Palatini}) \, ,
\end{equation}
where $\kappa = \sqrt{1 + 32 \xi}$, and in the limit $\xi \gg 1$ they reduce to~(\ref{cmbpalatini}).

\begin{figure}[h!]
    \centering
    \includegraphics[width=\textwidth]{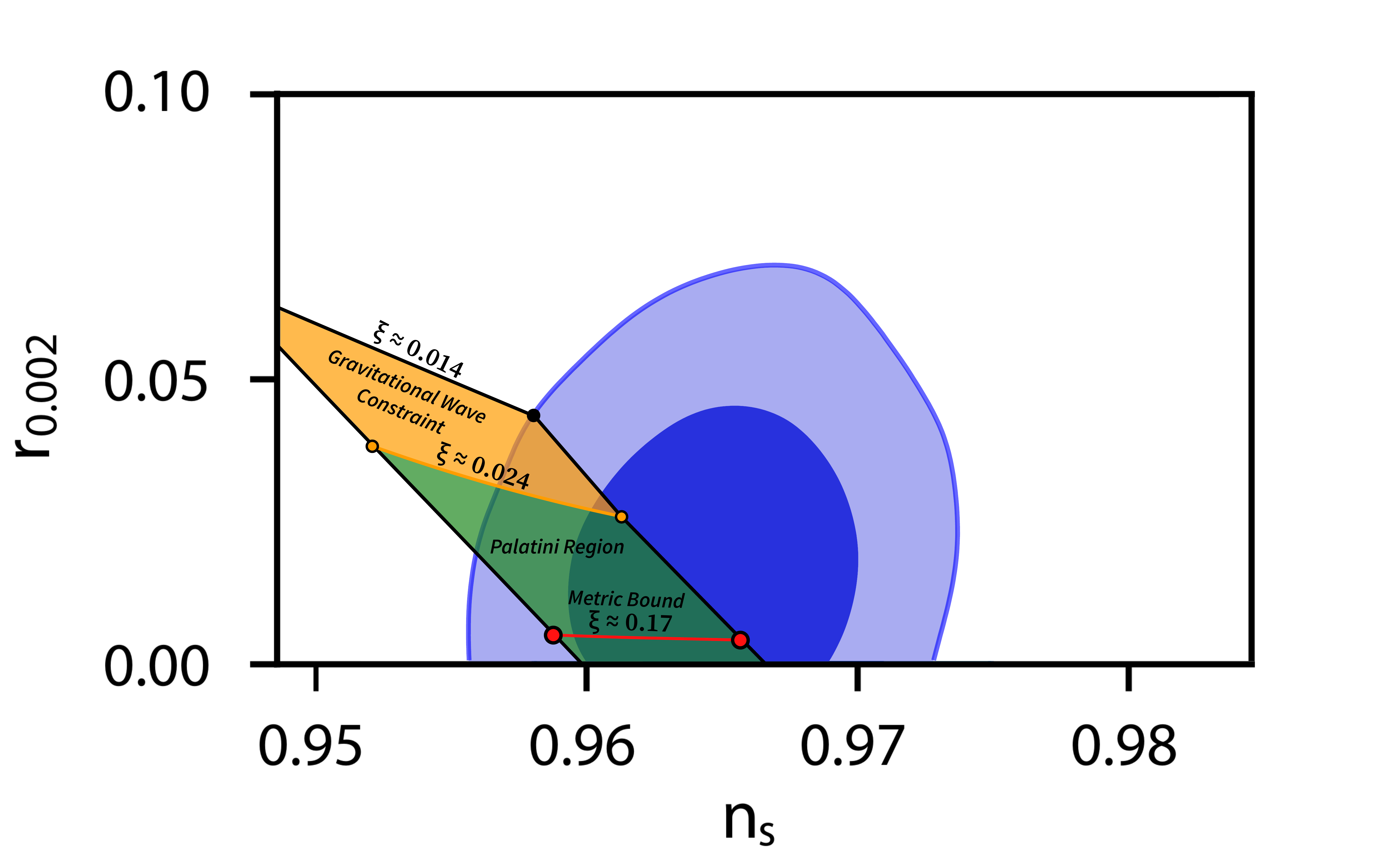}
    \caption{\textit{The cosmological observables $n_s$ and $r$ for the quartic potential in the Palatini formalism~(\ref{potpalatini}) between $N_* = 50$ (left) and $N_* = 60$ (right). The upper pair of black dots corresponds to a minimum value of $\xi_{\rm{min}} \simeq 0.014$ with $r \sim 0.05$ when $N_* = 60$. The region shaded orange corresponds to the parameter space excluded by the gravitational wave constraint~(\ref{gwcons2}), and the orange line indicates the maximum value of $\xi \simeq 0.024$ in this region. The region shaded green corresponds to the cosmological observables~(\ref{cmbpalatinifull}) predicted by the quartic potential~(\ref{potpalatini}) that are not excluded by the gravitational wave constraint. The red line indicates the CMB observables in the metric formalism (in the Palatini formalism this bound corresponds to a value of $\xi \simeq 0.17$), and the region below this line is predicted only in the Palatini theory of gravity. The blue shadings correspond to the $68 \%$ and $95 \%$ confidence levels from Planck data combined with BICEP2/Keck results~\cite{Aghanim:2018eyx}.}}
    \label{fig:planckdata}
\end{figure}

We show in Fig.~\ref{fig:planckdata} the principal CMB observables predicted by Palatini formalism~(\ref{cmbpalatinifull}) in the $(n_s, r)$ plane, together with the Planck data combined with BICEP2/Keck results~\cite{Aghanim:2018eyx}.

\section{Quintessential Inflation}
\label{sec:quintessence}
We begin this Section by recalling some of the features of the original model of \textit{quintessential inflation}, introduced by Peebles and Vilenkin~\cite{Peebles:1998qn}. The quintessential inflation potential is given by:
\renewcommand{\arraystretch}{2.0}
\begin{equation}
    \label{eq:quint1}
    V(\phi)=\left\{\begin{array}{ll}
    \dfrac{\lambda}{4}\left(\phi^{4}+M^{4}\right) & \text { for } \quad \phi<0 \, ,\\
    \dfrac{\lambda M^{4}}{4(1 + (\phi/M)^4)} & \text { for } \quad \phi \geq 0 \, ,
    \end{array}\right. 
\end{equation}
where $M \lesssim M_P$ is the
constant energy parameter that is responsible
for the present-day dark energy density. It is assumed that the inflaton field
starts rolling at $\phi \ll -1$ and inflation ends at $\phi \sim -1$. Following the inflationary epoch, most of the potential energy has been converted into the kinetic energy of the inflaton field and the Universe enters a period of kination. From the Klein-Gordon equation of motion, $\ddot{\phi} + 3 H \dot{\phi} \simeq 0$, we find that during kination, $\dot{\phi} \sim a^{-3}$, and the energy density of the inflaton scales as $\rho_{\phi} \simeq \dot{\phi}^2/2 \sim a^{-6}$. The key point is that $\rho_{\phi}$ decreases more rapidly than the radiation energy density, which scales as $\rho_r \sim a^{-4}$, and assuming that the potential energy density remains negligible, the energy density of the particles produced at the end of inflationary epoch eventually dominates, and reheating occurs.

When $\phi \gg 1$, the evolution of the inflaton field is governed by the inverse power-law (IPL) potential, which has a tracker solution~\cite{Zlatev:1998tr, Steinhardt:1999nw}. However, the inverse quartic potential $V \sim \phi^{-4}$ is too steep to follow the tracker solution, and instead the inflaton field rolls down its potential and freezes at the field value, $\phi_f$. In this case, the present-day dark energy density is explained by the potential energy density $V(\phi_f) \sim 10^{-120}$. 

One may consider also a more general inverse power-law quintessence model, given by
\begin{equation}
    \label{eq:quintgen}
    V(\phi) = \frac{\lambda M^4}{4(1+(\phi/M)^{\alpha})} \quad \text { for } \quad \phi \geq 0 \, ,
\end{equation}
where $\alpha > 0$. In freezing dark energy models~\cite{Ratra:1987rm}, the present-day dark energy density corresponds to $V(\phi_f) \sim \lambda M^{4 + \alpha}/\phi_{f}^{\alpha}$, and models with large values of $\alpha$ would exhibit a tracking behavior.\footnote{Potentials with tracking solutions must satisfy the tracking criterion $
\Gamma \equiv V^{\prime \prime} V/V^{\prime 2} \geq 1
$, and for inverse power-law potential~(\ref{eq:quintgen}), we find $\Gamma = (\alpha + 1)/\alpha$. In general, quintessence models with $\Gamma \simeq 1$ lead to a better tracking behavior, and for inverse power-law models, $\Gamma \rightarrow 1$ as $\alpha \rightarrow \infty$. For a more detailed discussion of tracking solutions in models of quintessence, see~\cite{Copeland:2006wr, Tsujikawa:2013fta, Bahamonde:2017ize}.} In this paper we focus on the case $\alpha = 4$, and different models of quintessence will be considered in future work.

As already mentioned in the previous Section, the quartic inflationary potential is excluded by the Planck data in theories with a \textit{minimal} coupling to gravity. Instead we consider a $\textit{non-minimal}$ model of quintessential inflation in Palatini formulation.\footnote{Simple models of power-law and inverse power-law inflation in Palatini formulation were studied in~\cite{Takahashi:2020car}. However, they were not discussed in the context of quintessence.} If we use the quintessential inflation potential~(\ref{eq:quint1}) with Eq.~(\ref{eq:potcan}), we obtain:
\begin{equation}
    \label{eq:palaquint}
    U(\chi)=\left\{\begin{array}{ll}
    \dfrac{\lambda\left(M^{4}+\dfrac{\sinh^4 (\sqrt{\xi} \chi)}{\xi^{2}}\right)}{4\left(1+\sinh^{2} (\sqrt{\xi} \chi)\right)^{2}} & \text { for } \quad \chi<0 \, , \\ 
    \dfrac{M^{8} \xi^{2} \lambda \operatorname{sech}^{4}(\sqrt{\xi} \chi)}{4\left(M^{4} \xi^{2}+\sinh^{4} (\sqrt{\xi} \chi)\right)} & \text { for } \quad \chi \geq 0 \, ,
\end{array}\right. 
\end{equation}
where during inflation the inflationary potential can be approximated by Eq.~(\ref{potpalatini}). After the inflationary epoch, when $\chi \gg 0$, the quintessential potential~(\ref{eq:palaquint}) can be approximated as
\begin{equation}
    \label{eq:palaquintapp}
    U(\chi) \simeq 4 M^8 \xi^2 \lambda \csch^{4}(2\sqrt{\xi} \chi) \, ,
\end{equation}
and we use this result in Section~\ref{sec:reh} to calculate the present-day dark energy density.

We conclude this Section by discussing briefly the $R^2$, or Starobinsky, model of inflation~\cite{Starobinsky:1980te}, that has been studied recently in the Palatini theory of gravity~\cite{Enckell:2018hmo, Antoniadis:2018ywb}. In this case, the $R^2$ term does not introduce a scalar degree of freedom, which in the metric formulation plays the role of the inflaton field. However, if we assume that in addition to the $R^2$ term, the Jordan frame action contains at least one dynamical scalar field coupled to gravity, then the inclusion of the $R^2$ term lowers the value of the tensor-to-scalar ratio, $r$, modifies the scalar potential form, and introduces higher-order kinetic terms. It would be interesting to study how the $R^2$ term affects the preheating mechanism and post-inflationary dynamics in quintessential inflation models, and we leave such considerations for future work.\footnote{For a related discussion, see~\cite{Giovannini:2019mgk}.}

\section{Preheating}
\label{sec:preheating}
In this Section, we review briefly the basic features of tachyonic preheating~\cite{Felder:2000hj, Felder:2001kt}. 
We follow the treatment of preheating in Palatini inflation, presented in Rubio and Tomberg~\cite{Rubio:2019ypq}, and apply it to the Peebles-Vilenkin model of quintessential inflation~(\ref{eq:palaquint}). However, here we do not discuss the preheating mechanism in metric formulation, which has been studied extensively in the context of Higgs inflation~\cite{GarciaBellido:2008ab, Bezrukov:2008ut}, and instead we focus on preheating dynamics in Palatini gravity.

It was shown in~\cite{Rubio:2019ypq} that the tachyonic production of scalar field fluctuations is extremely efficient and that the energy density of field fluctuations becomes comparable to the background energy density after a few oscillations of the inflaton field. Therefore, it can be considered as a possible reheating mechanism for non-oscillatory models of quintessential inflation. 

In order to study particle production during preheating, one can write the equation of motion for the scalar field fluctuations, $\psi_k$, in the Fourier space~\cite{Kofman:1997yn}:
\begin{equation}
\label{eq:pre1}
    \ddot{\psi}_k + 3H\dot{\psi}_k + \left(\frac{k^2}{a^2} + m^2_{\rm{eff}}(\chi) \right) \psi_k  =  0,
\end{equation}
which characterizes the \textit{self-resonance} of the inflaton field, where the effective mass term is defined as:
\begin{equation}
\label{eq:meff1}
    m_{\rm{eff}}^2(\chi) \equiv \frac{\partial^2 U(\chi)}{\partial \chi^2} = U''(\chi).
\end{equation}
In Eq.~(\ref{eq:pre1}) the gravitational particle production from the coupling to metric perturbations, which is subdominant compared to the particle production arising from a tachyonic instability, has been neglected. For more details related to the coupling to metric perturbations, see, e.g.,~\cite{Kaiser:2012ak}, and for a recent discussion of gravitational preheating in the context of Palatini gravity, see~\cite{Karam:2020rpa}.

To understand the behavior of Eq.~(\ref{eq:pre1}) in an expanding Universe, it is convenient to rescale the scalar field fluctuations $\psi_k(t)$ and introduce the function $\Psi_k(t) \equiv a^{3/2}(t) \psi_k(t)$. Then Eq.~(\ref{eq:pre1}) can be expressed in a simple form:
\begin{equation}
    \label{eq:pre2}
    \ddot{\Psi}_k +  \omega_k^2 \, \Psi_k = 0,
\end{equation}
with a time-dependent angular frequency,
\begin{equation}
    \label{eq:freq1}
    \omega_k^2 = \frac{k^2}{a^2} + U''(\chi),
\end{equation}
where the subdominant contribution $\Delta \equiv - \frac{3}{4} \left(\frac{\dot{a}}{a} \right)^2 - \frac{3}{2} \frac{\ddot{a}}{a}$ responsible for the gravitational particle production in an expanding Universe has been neglected. The mode equation~(\ref{eq:pre2}) describes a simple harmonic oscillator, with solutions $\Psi \sim e^{i \, t \sqrt{k^2/a^2 +  U''(\chi) }}$ that grow exponentially when the time-dependent angular frequency is negative. 
Therefore, from Eq.~(\ref{eq:freq1}), one can readily see that the tachyonic preheating occurs when
\begin{equation}
    \label{eq:conds1}
    \omega_k^2 < 0, \qquad \text{or} \qquad k^2 < - a^2 \, U''(\chi).
\end{equation}

The non-perturbative particle production during preheating stage can be characterized by the particle occupation number, $n_k$, which is an adiabatic invariant, and is given by~\cite{Kofman:1997yn}:
\begin{equation}
    \label{eq:occnum}
    n_k \; = \; \frac{\omega_k}{2} \left(\frac{|\dot{\Psi}_k|^2}{\omega_k^2} + |\Psi_k|^2 \right) - \frac{1}{2},
\end{equation}
and this definition is valid only for $\omega_k^2 > 0$. However, to use this 
expression in the tachyonic regime, one can instead define the angular frequency as $\omega_k = \sqrt{k^2/a^2 + \abs{U''(\chi)}}$ whenever $U''(\chi) < 0$. It is important to note that $n_k$ should not be treated as the particle occupation number during the tachyonic regime, and it can only be interpreted as such when $U''(\chi) > 0$. 
A more detailed discussion on the interpretation of $n_k$ in the tachyonic regime is presented in~\cite{Felder:2001kt}.

Since the particle occupation number $n_k$ is ill-defined in the tachyonic regime, it is more suitable to consider the energy density of scalar field fluctuations, given by
\begin{equation}
    \label{eq:endenpsi}
    \rho_{\Psi} \; = \;  \int_{0}^{k_{\rm{max}}} \frac{d^3k}{(2 \pi)^3} \frac{1}{2} \left[|\dot{\Psi}_k|^2 + \left(\frac{k^2}{a^2} + U''(\chi) \right)|\Psi_k|^2 \right],
\end{equation}
where the integration range $0 < k < {k_{\rm{max}}}$ represents the sub-Hubble modes in the tachyonic regime when $\omega_k^2 < 0$. We have used this result to calculate numerically the energy density of the created particles for the quintessential inflation potential~(\ref{eq:palaquint}) and we discuss it below.

It should be noted that we neglect the regularization for the tachyonic modes because they grow exponentially and their contribution to the total energy density is significantly larger than the vacuum energy contribution. However, momentum modes with $k > k_{\rm{max}}$ can be regularized but their contribution to the total energy density is subdominant, and it can be neglected by introducing a UV cut-off $k_{\rm{max}}$. For more details on cut-off regularization of the tachyonic modes in Palatini inflation, see~\cite{Rubio:2019ypq}.

Since the scalar field fluctuations are important only for sub-Hubble scales, the initial state is given by the Bunch-Davies vacuum~\cite{Birrell:1982ix}:
\begin{equation}
    \label{eq:bunchdavies}
    \Psi_{k}(t) = \frac{e^{-i \omega_k t}}{\sqrt{2 \omega_k}}, \qquad \dot{\Psi}_k(t) = -i \sqrt{\frac{\omega_k}{2}}e^{-i \omega_k t} \, ,
\end{equation}
and the scalar field fluctuations experience a net exponential growth, which is described by
\begin{equation}
    \label{eq:expgrowth}
    \Psi_k \sim \exp (\mu_k t),
\end{equation}
where $\mu_k$ is the growth index. 

\begin{figure}[h!]
    \centering
    \includegraphics[width=0.84\textwidth]{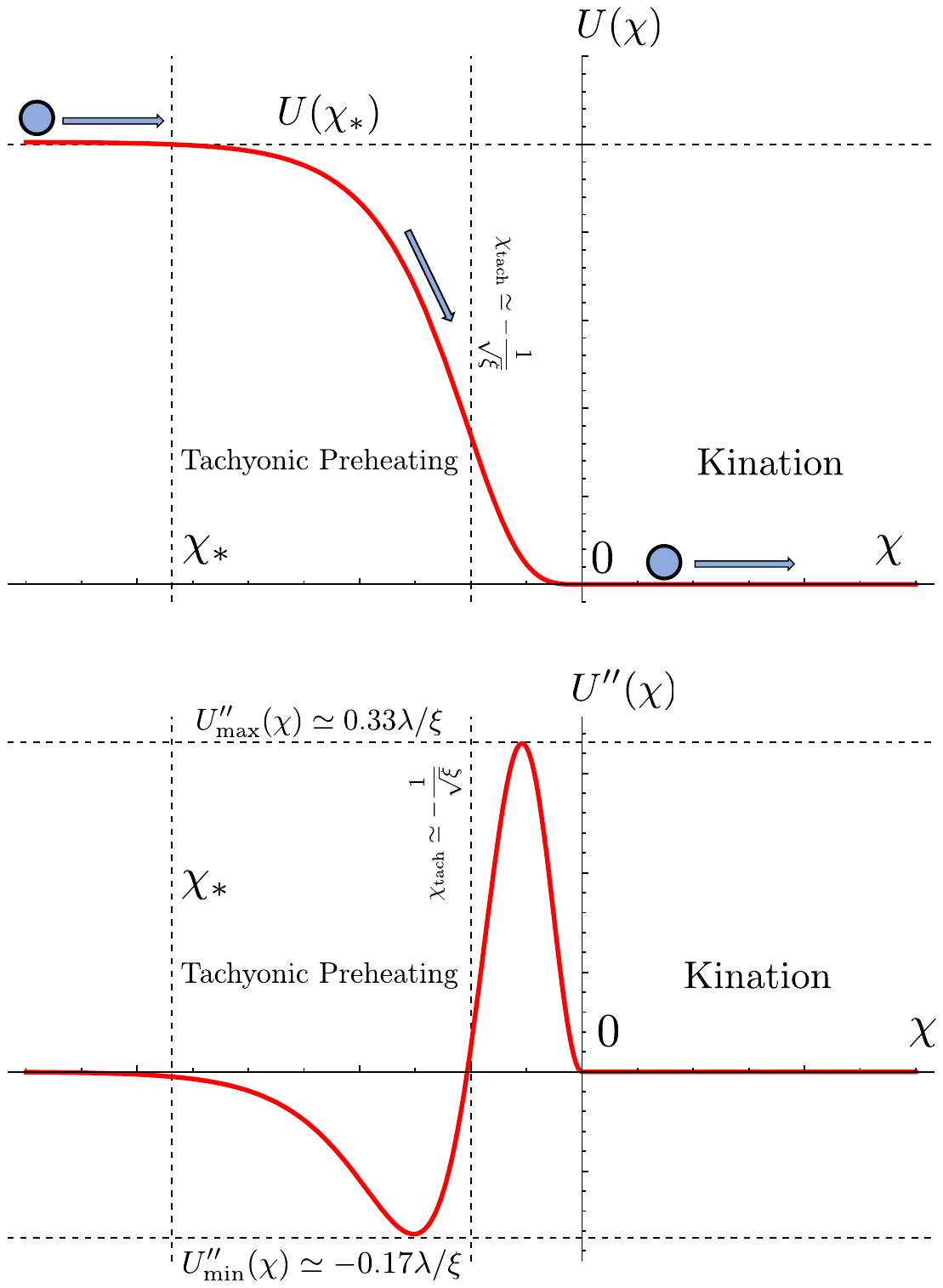}
    \caption{\textit{Top:  Quintessential inflation potential~(\ref{eq:palaquint}) as a function of $\chi$. The inflaton field slowly rolls down the potential and crosses the tachyonic region ($\omega_k^2 < 0$) where a
    very efficient preheating process occurs due to exponential growth of scalar field fluctuations. When the inflaton field crosses the zero point the Universe enters a kination phase. Bottom: The effective mass term~(\ref{eq:meff2}) as a function of $\chi$. The minimum is located at $U''_{\rm{min}(\chi)} \simeq -0.17 \lambda/\xi$ and the tachyonic preheating mechanism ends at $\chi_{\rm{end}} \simeq - 1/\sqrt{\xi}$ and $U''(\chi_{\rm{tach}}) = 0$.}}
    \label{fig:quintpot}
\end{figure}

Next, we study the tachyonic preheating in quintessential inflation. We find that the effective mass term for the inflationary potential~(\ref{eq:palaquint}) is given by
\begin{equation}
    \label{eq:meff2}
    U''(\chi) \; \simeq \; \frac{\lambda}{\xi} \left[4 - \cosh(2 \sqrt{\xi} \, \chi) \right] \sech^4(\sqrt{\xi} \, \chi) \tanh^2(\sqrt{\xi} \, \chi) \quad \rm{for} \quad \chi < 0 \, , 
\end{equation}
where we have omitted the constant energy parameter $M$ which is negligible during inflation, and the effective mass term $U''(\chi)$ is always positive for $ \chi \geq 0$. From Eq.~(\ref{eq:meff2}), we find that the maximum momentum value, $k_{\rm{max}}$, in the tachyonic regime ($\omega_k^2 < 0$) is given by
\begin{equation}
    k_{\rm{max}} = a \sqrt{\abs{U''_{\rm{min}} (\chi)}} \simeq 0.4 \, a \sqrt{\frac{\lambda}{\xi}},
\end{equation}
where $U''_{\rm{min}}(\chi)$ is the minimum value of the effective mass term, and tachyonic preheating ends when 
\begin{equation}
    \chi_{\rm{tach}} \simeq -\frac{\log(4 + \sqrt{15})}{2 \sqrt{\xi}}, \qquad U''(\chi_{\rm{tach}}) = 0,
\end{equation}
which agrees with the results in~\cite{Rubio:2019ypq}. Therefore, the tachyonic particle production occurs in the interval $\chi < \chi_{\rm{tach}}$. We show in Fig.~\ref{fig:quintpot} the quintessential inflation potential~(\ref{eq:palaquint}) and its corresponding effective mass term~(\ref{eq:meff2}) as a function of $\chi$.

After the inflaton field crosses the tachyonic region, the particle production
occurs via \textit{parametric resonance}~\cite{Kofman:1994rk, Shtanov:1994ce, Kofman:1997yn, Greene:1997fu}. However, this mechanism is very inefficient and we can safely neglect any additional particle production.

Further, we introduce the following energy density ratio:
\begin{equation}
    \label{eq:enratio}
    \Theta (t) = \frac{\rho_{\Psi}(t)}{\rho_{\phi}(t)} = \frac{1}{3H^2} \int_{0}^{k_{\rm{max}}} \frac{d^3k}{(2 \pi)^3} \frac{1}{2} \left[|\dot{\Psi}_k|^2 + \left(\frac{k^2}{a^2} + U''(\chi) \right)|\Psi_k|^2 \right] \, ,
\end{equation}
and we use this result to calculate the particle production numerically. It should be noted that in the case of non-oscillatory models, the backreaction effects are insignificant because the ratio $\Theta(t) \ll 0.1$. This bound is motivated by the lattice simulations of preheating~\cite{Enqvist:2015sua, Repond:2016sol}.

We assume that at the end of tachyonic preheating the scalar field fluctuations are converted into radiation. However, at the point $\chi_{\rm{tach}}$ the potential energy contribution to the total energy density is not negligible, and instead we assume that the Universe enters a period of kination at $t_0$, where $\chi(t_0) = 0$ is the point when the inflaton crosses the origin. We evaluate numerically the energy density ratio $\Theta(t_0)$~(\ref{eq:enratio}) for $N_* = 50$ and $60$. 
The minimum value of the coupling constant $\xi_{\rm{min}} \simeq 0.014$, which corresponds to $r \simeq 0.05$ when $N_* = 60$, is constrained by the Planck data and is shown in Fig.~\ref{fig:planckdata}. We summarize the key results in Table~\ref{tab:partheta} below.

\begin{table}[h!]
\centering
\begin{tabular}{|c|c|c|c|c|c|c|c|}
\hline
\multicolumn{4}{|c|}{\textbf{$N_* = 50$}}                      & \multicolumn{4}{c|}{\textbf{$N_* = 60$}}  \\ \hhline{|=|=|=|=|=|=|=|=|}
$\xi$  & $\lambda$ & $\Theta(t_0)$       & $t_0 \, (M_P^{-1})$               & $\xi$ & $\lambda$ & $\Theta(t_0)$ & $t_0 \, (M_P^{-1})$ \\ \hhline{|=|=|=|=|=|=|=|=|}
$0.01$ &      $1 \times 10^{-12}$     & $2 \times 10^{-16}$ & $5.3 \times 10^{6}$ &   $0.01$    &        $7 \times 10^{-13}$   & $3 \times 10^{-16}$              &  $6.9 \times 10^{6}$    \\ \hline
$0.05$ &      $5 \times 10^{-12}$     & $3 \times 10^{-13}$ & $5.7 \times 10^{6}$ &   $0.05$    &          $3 \times 10^{-12}$ &  $5 \times 10^{-13}$             &  $7.8 \times 10^{6}$     \\ \hline
$0.1$  &      $1 \times 10^{-11}$     & $1 \times 10^{-11}$ & $6.9 \times 10^{6}$ &   $0.1$    &          $7 \times 10^{-12}$ &    $1 \times 10^{-11}$           & $9.7 \times 10^{6}$      \\ \hline
$0.5$  &     $5 \times 10^{-11}$      & $1 \times 10^{-9}$  & $1.3 \times 10^{7}$ &  $0.5$     &         $3 \times 10^{-11}$  & $2 \times 10^{-9}$              &   $1.9 \times 10^{7}$    \\ \hline
$1.0$  &      $1 \times 10^{-10}$     & $7 \times 10^{-9}$  & $1.9 \times 10^{7}$ &  $1.0$     &        $7 \times 10^{-11}$   &        $7 \times 10^{-9}$       & $2.6 \times 10^{7}$      \\ \hline
\end{tabular}
\caption{\textit{Numerical results of tachyonic preheating for different parameters $\xi$ and $\lambda$ for $N_* = 50$ and $N_* = 60$. The energy density ratio $\Theta(t_0)$ is evaluated at time $t_0$ (in the units of $M_P^{-1}$) when the inflaton field crosses the zero point, $\chi(t_0) = 0$, and the Universe enters a kination phase.}}
\label{tab:partheta}
\end{table}

\section{Reheating and Gravitational Wave Constraint}
\label{sec:reh}
As discussed in Section~\ref{sec:slowroll}, in the slow-roll approximation the end of inflation is defined by $\epsilon(\chi_{\rm{end}}) = 1$, where from Eq.~(\ref{eq:slowrollpar}) we find that the field value at this point is given by:
\begin{equation}
    \label{eq:chiend}
    \chi_{\rm{end}} \simeq -\frac{\arcsinh{4\sqrt{2 \, \xi} }}{2 \sqrt{\xi}}.
\end{equation}
One can rewrite the slow-roll parameter $\epsilon$ as
\begin{equation}
    \label{eq:slowroll2}
    \epsilon = \frac{3}{2}(1 + w),
\end{equation}
where $w \equiv p/\rho$ is the equation-of-state parameter. When inflation ends, $w = -1/3$, which implies that at the end of inflation
\begin{equation}
    \label{eq:endcond}
    \dot{\chi}_{\rm{end}}^2 = U(\chi_{\rm{end}}),
\end{equation}
and at this point we cannot neglect the potential energy contribution to the total energy density of the inflaton, $\rho_{\chi}$.
Therefore, we do not assume that the transition from inflation to kination occurs instantly, and instead we assume that we enter a period of kination at time $t_0$, when the field value is $\chi(t_0) = 0$ and $\dot{\chi}(t_0)^2/2 \gg V$. For the numerical
results of the evolution from inflation to the present day in quintessential inflation where the transition from inflation to kination is non-instantaneous, see~\cite{Ellis:2020krl}.

During kination, the background field dynamics are described by the equations of motion:
\begin{equation}
    \label{eq:motback}
    \ddot{\chi} + 3 H \dot{\chi} \simeq 0, \qquad 3H^2 \simeq \frac{1}{2} \dot{\chi}^2,
\end{equation}
and in this regime, the inflaton energy density is given by $\rho_{\chi} \simeq \frac{\dot{\chi}^2}{2} \sim a^{-6}$, where during kination $a \sim t^{1/3}$ and the equation-of-state parameter is $w = 1$. If we use the expression~(\ref{eq:motback}) for the kination-dominated background, we obtain
\begin{equation}
    \label{eq:kinvelo}
    ~~~~\qquad \dot{\chi}(t)  = \sqrt{\frac{2}{3}} \, \frac{1}{t}, \quad {\rm{for}} \quad t_0 < t < t_{\rm{RH}} \, ,
\end{equation}    
and integrating this expression, we find
    \begin{equation}
    \label{eq:kinfield}
    \hspace{-2mm} \chi(t)  =  \sqrt{\frac{2}{3}} \log \left(\frac{t}{t_0} \right), \quad {\rm{for}} \quad t_0 < t < t_{\rm{RH}} \, ,
\end{equation}
where $t_{\rm{RH}}$ is the time when reheating occurs and we have used the condition
$\chi(t_0) = 0$.

In the previous Section, we have found the numerical values of the energy density ratio $\Theta(t_0)$ at time $t_0$ when the Universe enters a period of kination (see Table~\ref{tab:partheta}), and these parameters can be related to the reheating temperature, $T_{\rm{RH}}$. As noted earlier, when the inflaton kinetic energy dominates the background energy density, $a \sim t^{1/3}$, and the Hubble parameter is approximately $H(t) \sim 1/3t$. Therefore, at time $t_0$ the radiation energy density is given by
\begin{equation}
    \label{eq:radt0}
    \rho_r(t_0) \simeq \frac{\Theta(t_0)}{3 \, t_0^2} ,
\end{equation}
and this corresponds to a maximum temperature
\begin{equation}
    \label{eq:tmax}
    T_{\rm{max}} = \left( \frac{30  \rho_r(t_0)}{\pi^2 g_*} \right)^{1/4},
\end{equation}
where $g_* = 106.75$ is the number of relativistic degrees of freedom, and we assumed that the radiation has been thermalized.

When the Universe is dominated by the inflaton kinetic energy, $\Theta(t) = \rho_r(t)/\rho_{\chi}(t) \sim a^{2} \sim t^{2/3}$, and if we 
define the reheating temperature as the temperature when $\rho_r(t_{\rm{RH}}) \simeq \rho_{\chi}(t_{\rm{RH}})$, we find
\begin{equation}
    \label{eq:endenreh}
    \Theta(t_{\rm{RH}}) = \frac{\rho_{r}(t_{\rm{RH}})}{\rho_{\chi}(t_{\rm{RH}})} = 1, \qquad \rho_r(t_{\rm{RH}}) = \frac{\Theta(t_0)^3}{3 \, t_{0}^2},
\end{equation}
and the reheating temperature
\begin{equation}
    \label{eq:rehtem}
    T_{\mathrm{RH}}=\left(\frac{30 \rho_{r}\left(t_{\mathrm{RH}}\right)}{\pi^{2} g_{*}}\right)^{1 / 4} = \left( \frac{10 \, \Theta(t_0)^3}{\pi^2 g_* t_0^2} \right)^{1/4} \, .
\end{equation}

After reheating, the Universe becomes dominated by the radiation energy density, which scales as $\rho_r \sim a^{-4} \sim t^{-2}$, and the Hubble expansion rate is given by $H \sim 1/2t$. If we use the equation of motion in the radiation background, we find
\begin{equation}
    \label{eq:velrad}
    \dot{\chi} = \sqrt{\frac{2}{3}} \frac{\sqrt{t_{\rm{RH}}}}{t^{3/2}} \quad {\rm{for}} \quad t > t_{\rm{RH}},
\end{equation}
and if we integrate this expression, we find 
\begin{equation}
    \label{eq:posrad}
    \chi(t) = \sqrt{\frac{2}{3}} \log(\frac{t_{\rm{RH}}}{t_0}) + 2 \sqrt{\frac{2}{3}} \left( 1 - \sqrt{\frac{t_{\rm{RH}}}{t}}\right) \quad {\rm{for}} \quad t > t_{\rm{RH}} \, ,
\end{equation}
where we have used Eq.~(\ref{eq:kinfield}) for the field value $\chi(t_{\rm{RH}})$. Therefore, the field freezes when $t \gg t_{\rm{RH}}$, and combining the expression~(\ref{eq:posrad}) with
$\Theta(t_{\rm{RH}}) = \Theta(t_0)(t_{\rm{RH}}/t_0)^{2/3}= 1$ leads to
\begin{equation}
    \label{eq:fieldfreeze}
    \chi_F \simeq \sqrt{\frac{2}{3}} \left(2 - \frac{3}{2} \log{\Theta(t_0)} \right).
\end{equation}
Next, if we use this expression with the quintessential potential expression~(\ref{eq:palaquintapp}), where $\chi_F \gg 0$, and assume that the quintessence remains frozen until the present-day, we find
\begin{equation}
    \label{eq:potfrozen}
    U(\chi_F) \simeq 4 M^8 \xi^2 \lambda \csch^{4}\left(\sqrt{\frac{2 \xi}{3}} (4 - 3 \log{\Theta(t_0)} ) \right) \, ,
\end{equation}
where $U(\chi_F) \sim 10^{-120}$ is responsible for the cosmological constant (dark energy). In order to find the value of the constant energy density parameter $M$ required for this model, we use Eq.~(\ref{eq:potfrozen}) and obtain
\begin{equation}
    \label{eq:mterm}
    M = \left(\frac{U(\chi_F)}{4 \lambda \xi^2} \right)^{1/8} \sqrt{\sinh(\sqrt{\frac{2 \xi}{3}}(4-3\log{\Theta}(t_0)))},
\end{equation}
where we assume that $M \lesssim M_P$. From this upper bound we find numerically that the maximum value of the coupling constant is given by $\xi_{\rm{max}} \simeq 2$, which corresponds to the tensor-to-scalar ratio $r \sim \mathcal{O} (10^{-4})$.

The key results for our model, including the maximum temperature $T_{\rm{max}}$~(\ref{eq:tmax}), the reheating temperature $T_{\rm{RH}}$~(\ref{eq:rehtem}), and the value of $M$~(\ref{eq:mterm}) for $N_* = 50$ and $60$ are summarized in Table~\ref{tab:partheta2} below. From these results, we see that reheating occurs at a temperature above that of Big Bang Nucleosynthesis, $T_{\rm{RH}} \gg T_{\rm{BBN}} \simeq 1 \, \rm{MeV}$.

\begin{table}[h!]
\begin{adjustbox}{width=\columnwidth,center}
\centering
\begin{tabular}{|c|c|c|c|c|c|c|c|c|c|}
\hline
\multicolumn{5}{|c|}{\textbf{$N_* = 50$}}                      & \multicolumn{5}{c|}{\textbf{$N_* = 60$}}  \\ \hhline{|=|=|=|=|=|=|=|=|=|=|}
$\xi$  & $\lambda$ & $M(M_P)$      & $T_{\rm{max}} \, (\rm{GeV})$    &$T_{\rm{RH}} \, (\rm{GeV})$ & $\xi$ & $\lambda$ & $M(M_P)$      & $T_{\rm{max}} \, (\rm{GeV})$    &$T_{\rm{RH}} \, (\rm{GeV})$  \\ \hhline{|=|=|=|=|=|=|=|=|=|=|}
$0.01$ &      $1 \times 10^{-12}$     & $6 \times 10^{-12}$ & $3.9 \times 10^{10}$ &   $4.1 \times 10^2$    &  $0.01$  & $7 \times 10^{-13}$  & $6 \times 10^{-12}$   & $3.8 \times 10^{8}$              &  $4.9 \times 10^{2}$    \\ \hline
$0.05$ &      $5 \times 10^{-12}$     & $1 \times 10^{-10}$ & $2.3 \times 10^{11}$ &  $9.6 \times 10^{4}$ &$0.05$ & $3 \times 10^{-12}$   &  $1 \times 10^{-10}$ &  $2.3 \times 10^{11}$ &  $1.2 \times 10^{5}$     \\ \hline
$0.1$  &      $1 \times 10^{-11}$     & $8 \times 10^{-10}$ & $5.1 \times 10^{11}$ & $1.2 \times 10^{6}$  & $0.1$    &       $7 \times 10^{-12}$ & $8 \times 10^{-10}$ &   $4.2 \times 10^{11}$           & $1.0 \times 10^{6}$      \\ \hline
$0.5$  &     $5 \times 10^{-11}$      & $3 \times 10^{-6}$  & $1.2 \times 10^{12}$ & $2.8 \times 10^{7}$ & $0.5$     &         $3 \times 10^{-11}$  & $2 \times 10^{-6}$& $1.1 \times 10^{12}$              &   $3.9 \times 10^{7}$    \\ \hline
$1.0$  &      $1 \times 10^{-10}$     & $5 \times 10^{-4}$  & $1.6 \times 10^{12}$ & $1.0 \times 10^{8}$ & $1.0$     &        $7 \times 10^{-11}$   &        $5 \times 10^{-4}$       & $1.4 \times 10^{12}$  &  $8.5 \times 10^7$  \\ \hline
\end{tabular}
\end{adjustbox}
\caption{\textit{Parameters of the quintessential inflationary potential~(\ref{eq:palaquint}) for the specific choices of $N_* = 50$ and $60$. The value of the constant energy density parameter $M$ is determined by Eq.~(\ref{eq:potfrozen}), and the maximum and reheating temperatures are given by the expressions~(\ref{eq:tmax}) and~(\ref{eq:rehtem}), respectively.}}
\label{tab:partheta2}
\end{table}

However, when the background energy density is dominated by the kinetic energy of the inflaton field, the primordial gravitational wave spectrum becomes blue-tilted. The energy density ratio of primordial gravitational waves relative to the critical energy density in a flat Universe can be expressed as~\cite{Sahni:1990tx}:
\begin{equation}
    \label{eq:gravwaves}
    \Omega_{\mathrm{GW}}=\frac{1}{\rho_{c}} \frac{d \rho_{\mathrm{GW}}}{d \ln k} \propto k^{2\left(\frac{3 w-1}{3 w+1}\right)} \propto a^{3 w-1} \, ,
\end{equation}
where $\rho_{c} \equiv 3 H_{0}^{2} M_{P}^{2} \simeq 1.05 \, h^{2} \times 10^{-5} \, \mathrm{GeV} \, \mathrm{cm}^{-3}$ is the
critical density of the Universe, $\rho_{\mathrm{GW}}$ is the energy density of the gravitational waves, and $k=a H$ is the momentum at the Hubble horizon crossing. The energy density fraction~(\ref{eq:gravwaves}) is an increasing function of $a$ for $w > 1/3$, and it scales as $\Omega_{\rm{GW}} \sim k \sim a^{2}$ when $w = 1$. Therefore, a prolonged period of kination leads to a blue-tilted primordial gravitational wave spectrum and, in turn, an excessive contribution to the total radiation energy density of the Universe. Consequently, this would change the effective number of neutrino species, $N_{\rm{eff}}$, and affect Big Bang Nucleosynthesis. 

We use the current upper bound on $N_{\rm{eff}} < 3.17$~\cite{Fields:2019pfx}, which sets a limit on the gravitational wave spectrum today~\cite{Maggiore:1999vm},
\begin{equation}
    \label{eq:gravcons}
    \Omega_{\mathrm{GW}}<\left(\frac{4}{11}\right)^{4 / 3} \frac{7}{8}\left(N_{\mathrm{eff}}-3\right) \Omega_{\gamma} \, ,
\end{equation}
where $\Omega_{\gamma}=\left(\pi^{2} / 15\right) T_{0}^{4} / \rho_{c} \simeq 2.47 \times 10^{-5} h^{-2}$ is the radiation energy density today, which is dominated by the energy in the CMB. In order to ensure that the additional contribution from the blue-tilted primordial gravitational wave spectrum does not affect Big Bang Nucleosynthesis, we impose the limit~\cite{Maggiore:1999vm}:
\begin{equation}
    \label{gwbound}
    \mathcal{I} \equiv h^2 \int_{k_{\rm{BBN}}}^{k_0} \Omega_{\rm{GW}} \, d \log{k} \lesssim 10^{-6},
\end{equation}
where the lower bound $k_{\rm{BBN}}$ and the upper bound $k_0$ correspond to the momentum mode at BBN and the beginning of kination-domination, respectively. When the Universe becomes radiation-dominated, $w = 1/3$, and from Eq.~(\ref{eq:gravwaves}) one can see that the gravitational wave spectrum scales as $\Omega_{\rm{GW}} \sim \rm{const.}$ Therefore, we can consider the momentum range $k_{\rm{RH}} \lesssim k \lesssim k_{0}$, where $k_{\rm{RH}}$ is the momentum mode at reheating, and neglect the momentum modes $k_{\rm{BBN}} \lesssim k \lesssim k_{\rm{RH}}$.

When the Universe is dominated by the inflaton kinetic energy density, the gravitational wave spectrum is given by~\cite{Giovannini:1999bh, Giovannini:1999qj}
\begin{equation}
    \label{eq:gravkin1}
    \Omega_{\mathrm{GW}} \simeq \varepsilon \, \Omega_{\gamma} h_{\mathrm{GW}}^{2}\left(\frac{k}{k_{\mathrm{RH}}}\right)\left[\ln \left(\frac{k}{k_0}\right)\right]^{2} \quad {\rm{for}} \quad k_{\rm{RH}} \lesssim k \lesssim k_{0} \, ,
\end{equation}
where $h_{\mathrm{GW}}^{2}=\frac{1}{8 \pi}\left(\frac{H\left(t_{0}\right)}{M_{P}}\right)^{2}$ is the amplitude of the gravitational waves and $\varepsilon=2 R_{i}\left(\frac{3.36}{g_{*}}\right)^{1 / 3}$, where $R_{i}=\frac{81}{32 \pi^{3}}$ is the energy density contribution from the massless scalar degrees of freedom. If we use the above expression with the bound~(\ref{gwbound}), we find
\begin{equation}
    \label{eq:gwboundsimp}
    \mathcal{I} \simeq 2 \varepsilon h_{\mathrm{GW}}^{2} \Omega_{\gamma} h^{2}\left(\frac{k_{0}}{k_{\mathrm{RH}}}\right).
\end{equation}
We can express the ratio $k_0/k_{\rm{RH}}$ as
\begin{equation}
    \frac{k_0}{k_{\rm{RH}}} = \frac{a(t_0)}{a(t_{\rm{RH}}) } \frac{H(t_0)}{H(t_{\rm{RH}})} = \left(\frac{t_{\rm{RH}}}{t_0} \right)^{2/3} = \frac{1}{\Theta(t_0)} \, ,
\end{equation}
and if we rewrite the gravitational wave amplitude as $h_{\rm{GW}}^2 = 1/72 \pi t_0^2$, we obtain
\begin{equation}
    \label{gwcons2}
    \mathcal{I} \simeq \frac{10^{-8}}{\Theta(t_0) \, t_0^2} \lesssim 10^{-6}, \quad {\text{or}} \quad \Theta(t_0) \, t_0^2 \gtrsim 0.01.
\end{equation}
We find numerically that this upper limit is satisfied for $\xi_{\rm{GW}} \gtrsim 0.024$, and the excluded region is shown in Fig.~\ref{fig:planckdata}.

\section{Results and Discussion}
\label{sec:results}
We have presented in this paper a model of quintessential inflation with a non-minimal coupling in the Palatini formulation of gravity. We focused our attention on the original model of quintessential inflation, where the inflationary epoch is characterized by a quartic potential and after inflation, the inflaton field rolls down the inverse power-law potential. Since the measurements of the CMB spectrum rule out the quartic potential, we have studied this model in Palatini theory with a small coupling to gravity, whose predictions agree with the most recent Planck data, and they are shown in Fig.~\ref{fig:planckdata}.

At the end of inflationary epoch, the inflaton field passes through a tachyonic region, which leads to a very efficient particle production via tachyonic preheating. We have performed a detailed numerical study of the preheating era and summarized the results in Table~\ref{tab:partheta}. Following inflation, the Universe becomes dominated by the kinetic energy of the inflaton and enters a period of kination until the background energy density becomes dominated by radiation and reheating occurs. We have found that in these models reheating occurs at a temperature $T_{\rm{RH}} \sim \mathcal{O}(10^3 - 10^8) \, \rm{GeV}$, which is significantly above the temperature of Big Bang Nucleosynthesis, $T_{\rm{BBN}} \sim 1 \, \rm{MeV}$, and the results are tabulated in Table~\ref{tab:partheta2}.

After reheating, the inflaton field rolls down the quintessence potential and eventually freezes at $\chi_F$. Since the field remains frozen until the present day, it plays the role of dark energy, which is given by $U(\chi_F) \sim 10^{-120}$. We have found that these quintessential inflation models are viable for small values of the coupling $0.024 \lesssim \xi \lesssim 2$, which correspond to the tensor-to-scalar ratio interval $10^{-4} \lesssim r \lesssim 0.03$. In contrast, the metric formulation predicts the tensor-to-scalar ratio between $0.003 \lesssim r \lesssim 0.005$, and the upcoming CMB measurements may be used to distinguish between metric and Palatini theories of gravity.

We conclude that this model of quintessential inflation is a promising alternative to conventional models of inflation, and calls for further study. It would be interesting to study different models of quintessence and the effects of the higher-order curvature terms on preheating mechanism. Finally, we have provided a concrete example of quintessential inflation that offers the possibility of relating directly the theory of inflation to different theories of gravity.

\subsection*{Acknowledgements}
\noindent
I would like to thank Keith Olive and Marcos Garc{\' i}a
for useful discussions and comments.


\newpage
\begin{thebibliography}{9}

\bibitem{Olive:1989nu}
K.~A.~Olive,
Phys. Rept. \textbf{190}, 307-403 (1990)

\bibitem{Linde:2005ht}
A.~D.~Linde,
Contemp. Concepts Phys. \textbf{5}, 1-362 (1990)
[arXiv:hep-th/0503203 [hep-th]].

\bibitem{Lyth:1998xn}
D.~H.~Lyth and A.~Riotto,
Phys. Rept. \textbf{314}, 1-146 (1999)
[arXiv:hep-ph/9807278 [hep-ph]].

\bibitem{Martin:2013tda}
J.~Martin, C.~Ringeval and V.~Vennin,
Phys. Dark Univ. \textbf{5-6}, 75-235 (2014)
[arXiv:1303.3787 [astro-ph.CO]].

\bibitem{Martin:2013nzq}
J.~Martin, C.~Ringeval, R.~Trotta and V.~Vennin,
JCAP \textbf{03}, 039 (2014)
doi:10.1088/1475-7516/2014/03/039
[arXiv:1312.3529 [astro-ph.CO]].

\bibitem{Martin:2015dha}
J.~Martin,
Astrophys. Space Sci. Proc. \textbf{45}, 41-134 (2016)
[arXiv:1502.05733 [astro-ph.CO]].

\bibitem{Dolgov:1982th}
A.~D.~Dolgov and A.~D.~Linde,
Phys. Lett. B \textbf{116}, 329 (1982)

\bibitem{Abbott:1982hn}
L.~F.~Abbott, E.~Farhi and M.~B.~Wise,
Phys. Lett. B \textbf{117}, 29 (1982)

\bibitem{Nanopoulos:1983up}
D.~V.~Nanopoulos, K.~A.~Olive and M.~Srednicki,
Phys. Lett. B \textbf{127}, 30-34 (1983)

\bibitem{Aghanim:2018eyx}
N.~Aghanim \textit{et al.} [Planck],
Astron. Astrophys. \textbf{641}, A6 (2020)
[arXiv:1807.06209 [astro-ph.CO]].

\bibitem{Akrami:2018odb}
Y.~Akrami \textit{et al.} [Planck],
Astron. Astrophys. \textbf{641}, A10 (2020)
[arXiv:1807.06211 [astro-ph.CO]].

\bibitem{Ade:2018gkx}
P.~A.~R.~Ade \textit{et al.} [BICEP2 and Keck Array],
Phys. Rev. Lett. \textbf{121}, 221301 (2018)
[arXiv:1810.05216 [astro-ph.CO]].

\bibitem{Matsumura:2013aja}
T.~Matsumura, Y.~Akiba, J.~Borrill, Y.~Chinone, M.~Dobbs, H.~Fuke, A.~Ghribi, M.~Hasegawa, K.~Hattori and M.~Hattori, \textit{et al.}
J. Low Temp. Phys. \textbf{176}, 733 (2014)
[arXiv:1311.2847 [astro-ph.IM]].

\bibitem{Ade:2018sbj}
P.~Ade \textit{et al.} [Simons Observatory],
JCAP \textbf{02}, 056 (2019)
[arXiv:1808.07445 [astro-ph.CO]].

\bibitem{Hanany:2019lle}
S.~Hanany \textit{et al.} [NASA PICO],
[arXiv:1902.10541 [astro-ph.IM]].

\bibitem{Abazajian:2016yjj}
K.~N.~Abazajian \textit{et al.} [CMB-S4],
[arXiv:1610.02743 [astro-ph.CO]].


\bibitem{Tenkanen:2020dge}
T.~Tenkanen,
Gen. Rel. Grav. \textbf{52}, no.4, 33 (2020)
[arXiv:2001.10135 [astro-ph.CO]].

\bibitem{Shimada:2018lnm}
K.~Shimada, K.~Aoki and K.~i.~Maeda,
Phys. Rev. D \textbf{99}, no.10, 104020 (2019)
[arXiv:1812.03420 [gr-qc]].

\bibitem{Bauer:2008zj}
F.~Bauer and D.~A.~Demir,
Phys. Lett. B \textbf{665}, 222-226 (2008)
[arXiv:0803.2664 [hep-ph]].

\bibitem{Bauer:2010jg}
F.~Bauer and D.~A.~Demir,
Phys. Lett. B \textbf{698}, 425-429 (2011)
[arXiv:1012.2900 [hep-ph]].

\bibitem{Tamanini:2010uq}
N.~Tamanini and C.~R.~Contaldi,
Phys. Rev. D \textbf{83}, 044018 (2011)
[arXiv:1010.0689 [gr-qc]].

\bibitem{Tenkanen:2017jih}
T.~Tenkanen,
JCAP \textbf{12}, 001 (2017)
[arXiv:1710.02758 [astro-ph.CO]].

\bibitem{Antoniadis:2018yfq}
I.~Antoniadis, A.~Karam, A.~Lykkas, T.~Pappas and K.~Tamvakis,
JCAP \textbf{03}, 005 (2019)
[arXiv:1812.00847 [gr-qc]].





\bibitem{Fu:2017iqg}
C.~Fu, P.~Wu and H.~Yu,
Phys. Rev. D \textbf{96}, no.10, 103542 (2017)
[arXiv:1801.04089 [gr-qc]].

\bibitem{Jarv:2017azx}
L.~Järv, A.~Racioppi and T.~Tenkanen,
Phys. Rev. D \textbf{97}, no.8, 083513 (2018)
[arXiv:1712.08471 [gr-qc]].

\bibitem{Racioppi:2018zoy}
A.~Racioppi,
Phys. Rev. D \textbf{97}, no.12, 123514 (2018)
[arXiv:1801.08810 [astro-ph.CO]].

\bibitem{Carrilho:2018ffi}
P.~Carrilho, D.~Mulryne, J.~Ronayne and T.~Tenkanen,
JCAP \textbf{06}, 032 (2018)
[arXiv:1804.10489 [astro-ph.CO]].

\bibitem{Enckell:2018hmo}
V.~M.~Enckell, K.~Enqvist, S.~Rasanen and L.~P.~Wahlman,
JCAP \textbf{02}, 022 (2019)
[arXiv:1810.05536 [gr-qc]].

\bibitem{Antoniadis:2018ywb}
I.~Antoniadis, A.~Karam, A.~Lykkas and K.~Tamvakis,
JCAP \textbf{11}, 028 (2018)
[arXiv:1810.10418 [gr-qc]].

\bibitem{Almeida:2018oid}
J.~P.~B.~Almeida, N.~Bernal, J.~Rubio and T.~Tenkanen,
JCAP \textbf{03}, 012 (2019)
[arXiv:1811.09640 [hep-ph]].

\bibitem{Takahashi:2018brt}
T.~Takahashi and T.~Tenkanen,
JCAP \textbf{04}, 035 (2019)
[arXiv:1812.08492 [astro-ph.CO]].

\bibitem{Jinno:2018jei}
R.~Jinno, K.~Kaneta, K.~y.~Oda and S.~C.~Park,
Phys. Lett. B \textbf{791}, 396-402 (2019)
[arXiv:1812.11077 [gr-qc]].

\bibitem{Edery:2019txq}
A.~Edery and Y.~Nakayama,
Phys. Rev. D \textbf{99}, no.12, 124018 (2019)
[arXiv:1902.07876 [hep-th]].

\bibitem{Giovannini:2019mgk}
M.~Giovannini,
Class. Quant. Grav. \textbf{36}, no.23, 235017 (2019)
[arXiv:1905.06182 [gr-qc]].

\bibitem{Tenkanen:2019wsd}
T.~Tenkanen,
Phys. Rev. D \textbf{101}, no.6, 063517 (2020)
[arXiv:1910.00521 [astro-ph.CO]].

\bibitem{Bostan:2019wsd}
N.~Bostan,
Commun. Theor. Phys. \textbf{72}, 085401 (2020)
[arXiv:1908.09674 [astro-ph.CO]].


\bibitem{Gialamas:2019nly}
I.~D.~Gialamas and A.~B.~Lahanas,
Phys. Rev. D \textbf{101}, no.8, 084007 (2020)
[arXiv:1911.11513 [gr-qc]].

\bibitem{Racioppi:2019jsp}
A.~Racioppi,
[arXiv:1912.10038 [hep-ph]].

\bibitem{Tenkanen:2020cvw}
T.~Tenkanen and E.~Tomberg,
JCAP \textbf{04}, 050 (2020)
[arXiv:2002.02420 [astro-ph.CO]].

\bibitem{Lloyd-Stubbs:2020pvx}
A.~Lloyd-Stubbs and J.~McDonald,
Phys. Rev. D \textbf{101}, no.12, 123515 (2020)
[arXiv:2002.08324 [hep-ph]].

\bibitem{Antoniadis:2020dfq}
I.~Antoniadis, A.~Lykkas and K.Tamvakis, 
JCAP \textbf{04}, no.04, 033 (2020)
[arXiv:2002.12681 [gr-qc]].

\bibitem{Ghilencea:2020piz}
D.~M.~Ghilencea,
[arXiv:2003.08516 [hep-th]].

\bibitem{Takahashi:2020car}
T.~Takahashi, T.~Tenkanen and S.~Yokoyama,
Phys. Rev. D \textbf{102}, no.4, 043524 (2020)
[arXiv:2003.10203 [astro-ph.CO]].

\bibitem{Das:2020kff}
N.~Das and S.~Panda,
[arXiv:2005.14054 [gr-qc]].

\bibitem{Jarv:2020qqm}
L.~Järv, A.~Karam, A.~Kozak, A.~Lykkas, A.~Racioppi and M.~Saal,
Phys. Rev. D \textbf{102}, no.4, 044029 (2020)
[arXiv:2005.14571 [gr-qc]].

\bibitem{Gialamas:2020snr}
I.~D.~Gialamas, A.~Karam and A.~Racioppi,
[arXiv:2006.09124 [gr-qc]].

\bibitem{Karam:2020rpa}
A.~Karam, M.~Raidal and E.~Tomberg,
[arXiv:2007.03484 [astro-ph.CO]].








\bibitem{Bezrukov:2007ep}
F.~L.~Bezrukov and M.~Shaposhnikov,
Phys. Lett. B \textbf{659}, 703-706 (2008)
[arXiv:0710.3755 [hep-th]].

\bibitem{Rubio:2018ogq}
J.~Rubio,
Front. Astron. Space Sci. \textbf{5}, 50 (2019)
[arXiv:1807.02376 [hep-ph]].



\bibitem{Rasanen:2017ivk}
S.~Rasanen and P.~Wahlman,
JCAP \textbf{11}, 047 (2017)
[arXiv:1709.07853 [astro-ph.CO]].

\bibitem{Racioppi:2017spw}
A.~Racioppi,
JCAP \textbf{12}, 041 (2017)
[arXiv:1710.04853 [astro-ph.CO]].

\bibitem{Markkanen:2017tun}
T.~Markkanen, T.~Tenkanen, V.~Vaskonen and H.~Veermäe,
JCAP \textbf{03}, 029 (2018)
[arXiv:1712.04874 [gr-qc]].

\bibitem{Enckell:2018kkc}
V.~M.~Enckell, K.~Enqvist, S.~Rasanen and E.~Tomberg,
JCAP \textbf{06}, 005 (2018)
[arXiv:1802.09299 [astro-ph.CO]].

\bibitem{Kannike:2018zwn}
K.~Kannike, A.~Kubarski, L.~Marzola and A.~Racioppi,
Phys. Lett. B \textbf{792}, 74-80 (2019)
[arXiv:1810.12689 [hep-ph]].

\bibitem{Rasanen:2018fom}
S.~Rasanen and E.~Tomberg,
JCAP \textbf{01}, 038 (2019)
[arXiv:1810.12608 [astro-ph.CO]].

\bibitem{Rasanen:2018ihz}
S.~Rasanen,
Open J. Astrophys. \textbf{2}, no.1, 1 (2019)
[arXiv:1811.09514 [gr-qc]].

\bibitem{Tenkanen:2019jiq}
T.~Tenkanen,
Phys. Rev. D \textbf{99}, no.6, 063528 (2019)
[arXiv:1901.01794 [astro-ph.CO]].

\bibitem{Rubio:2019ypq}
J.~Rubio and E.~S.~Tomberg,
JCAP \textbf{04}, 021 (2019)
[arXiv:1902.10148 [hep-ph]].

\bibitem{Jinno:2019und}
R.~Jinno, M.~Kubota, K.~y.~Oda and S.~C.~Park,
JCAP \textbf{03}, 063 (2020)
[arXiv:1904.05699 [hep-ph]].

\bibitem{Tenkanen:2019xzn}
T.~Tenkanen and L.~Visinelli,
JCAP \textbf{08}, 033 (2019)
[arXiv:1906.11837 [astro-ph.CO]].

\bibitem{Shaposhnikov:2020geh}
M.~Shaposhnikov, A.~Shkerin and S.~Zell,
[arXiv:2001.09088 [hep-th]].

\bibitem{Shaposhnikov:2020fdv}
M.~Shaposhnikov, A.~Shkerin and S.~Zell,
JCAP \textbf{07}, 064 (2020)
[arXiv:2002.07105 [hep-ph]].

\bibitem{McDonald:2020lpz}
J.~McDonald,
[arXiv:2007.04111 [hep-ph]].

\bibitem{Gialamas:2020vto}
I.~D.~Gialamas, A.~Karam, A.~Lykkas and T.~D.~Pappas,
Phys. Rev. D \textbf{102}, 063522 (2020)
[arXiv:2008.06371 [gr-qc]].


\bibitem{Peebles:1998qn}
P.~J.~E.~Peebles and A.~Vilenkin,
Phys. Rev. D \textbf{59}, 063505 (1999)
[arXiv:astro-ph/9810509 [astro-ph]].

\bibitem{Parker:1968mv}
L.~Parker,
Phys. Rev. Lett. \textbf{21}, 562-564 (1968)

\bibitem{Grib:1976pw}
A.~A.~Grib, S.~G.~Mamaev and V.~M.~Mostepanenko,
Gen. Rel. Grav. \textbf{7}, 535-547 (1976)

\bibitem{Zeldovich:1977vgo}
Y.~B.~Zel'dovich and A.~A.~Starobinsky,
JETP Lett. \textbf{26}, no.5, 252 (1977)

\bibitem{Ford:1986sy}
L.~H.~Ford,
Phys. Rev. D \textbf{35}, 2955 (1987)

\bibitem{Hashiba:2018iff}
S.~Hashiba and J.~Yokoyama,
JCAP \textbf{01}, 028 (2019)
[arXiv:1809.05410 [gr-qc]].
  
\bibitem{Felder:1999pv}
G.~N.~Felder, L.~Kofman and A.~D.~Linde,
Phys. Rev. D \textbf{60}, 103505 (1999)
[arXiv:hep-ph/9903350 [hep-ph]].
 
\bibitem{Felder:1998vq}
G.~N.~Felder, L.~Kofman and A.~D.~Linde,
Phys. Rev. D \textbf{59}, 123523 (1999)
[arXiv:hep-ph/9812289 [hep-ph]].

\bibitem{Kofman:1994rk}
L.~Kofman, A.~D.~Linde and A.~A.~Starobinsky,
Phys. Rev. Lett. \textbf{73}, 3195-3198 (1994)
[arXiv:hep-th/9405187 [hep-th]].

\bibitem{Shtanov:1994ce}
Y.~Shtanov, J.~H.~Traschen and R.~H.~Brandenberger,
Phys. Rev. D \textbf{51}, 5438-5455 (1995)
[arXiv:hep-ph/9407247 [hep-ph]].

\bibitem{Kofman:1997yn}
L.~Kofman, A.~D.~Linde and A.~A.~Starobinsky,
Phys. Rev. D \textbf{56}, 3258-3295 (1997)
[arXiv:hep-ph/9704452 [hep-ph]].

\bibitem{Greene:1997fu}
P.~B.~Greene, L.~Kofman, A.~D.~Linde and A.~A.~Starobinsky,
Phys. Rev. D \textbf{56}, 6175-6192 (1997)
[arXiv:hep-ph/9705347 [hep-ph]].


\bibitem{Felder:2000hj}
G.~N.~Felder, J.~Garcia-Bellido, P.~B.~Greene, L.~Kofman, A.~D.~Linde and I.~Tkachev,
Phys. Rev. Lett. \textbf{87}, 011601 (2001)
doi:10.1103/PhysRevLett.87.011601
[arXiv:hep-ph/0012142 [hep-ph]].

\bibitem{Felder:2001kt}
G.~N.~Felder, L.~Kofman and A.~D.~Linde,
Phys. Rev. D \textbf{64}, 123517 (2001)
[arXiv:hep-th/0106179 [hep-th]].

\bibitem{Linde:2011nh}
A.~Linde, M.~Noorbala and A.~Westphal,
JCAP \textbf{03}, 013 (2011)
[arXiv:1101.2652 [hep-th]].

\bibitem{Kallosh:2013hoa}
R.~Kallosh and A.~Linde,
JCAP \textbf{07}, 002 (2013)
[arXiv:1306.5220 [hep-th]].

\bibitem{Kallosh:2013pby}
R.~Kallosh and A.~Linde,
JCAP \textbf{06}, 027 (2013)
[arXiv:1306.3211 [hep-th]].
  
\bibitem{GarciaBellido:2008ab}
J.~Garcia-Bellido, D.~G.~Figueroa and J.~Rubio,
Phys. Rev. D \textbf{79}, 063531 (2009)
[arXiv:0812.4624 [hep-ph]].
  
\bibitem{Liddle:2003as}
A.~R.~Liddle and S.~M.~Leach,
Phys. Rev. D \textbf{68}, 103503 (2003)
[arXiv:astro-ph/0305263 [astro-ph]].

\bibitem{Martin:2010kz}
J.~Martin and C.~Ringeval,
Phys. Rev. D \textbf{82}, 023511 (2010)
[arXiv:1004.5525 [astro-ph.CO]].
 
\bibitem{Kaiser:2012ak}
D.~I.~Kaiser, E.~A.~Mazenc and E.~I.~Sfakianakis,
Phys. Rev. D \textbf{87}, 064004 (2013)
[arXiv:1210.7487 [astro-ph.CO]].

\bibitem{Zlatev:1998tr}
I.~Zlatev, L.~M.~Wang and P.~J.~Steinhardt,
Phys. Rev. Lett. \textbf{82}, 896-899 (1999)
[arXiv:astro-ph/9807002 [astro-ph]].

\bibitem{Steinhardt:1999nw}
P.~J.~Steinhardt, L.~M.~Wang and I.~Zlatev,
Phys. Rev. D \textbf{59}, 123504 (1999)
[arXiv:astro-ph/9812313 [astro-ph]].

\bibitem{Ratra:1987rm}
B.~Ratra and P.~J.~E.~Peebles,
Phys. Rev. D \textbf{37}, 3406 (1988)

\bibitem{Copeland:2006wr}
E.~J.~Copeland, M.~Sami and S.~Tsujikawa,
Int. J. Mod. Phys. D \textbf{15}, 1753-1936 (2006)
doi:10.1142/S021827180600942X
[arXiv:hep-th/0603057 [hep-th]].

\bibitem{Tsujikawa:2013fta}
S.~Tsujikawa,
Class. Quant. Grav. \textbf{30}, 214003 (2013)
[arXiv:1304.1961 [gr-qc]].


\bibitem{Bahamonde:2017ize}
S.~Bahamonde, C.~G.~B\"ohmer, S.~Carloni, E.~J.~Copeland, W.~Fang and N.~Tamanini,
Phys. Rept. \textbf{775-777}, 1-122 (2018)
[arXiv:1712.03107 [gr-qc]].

\bibitem{Starobinsky:1980te}
A.~A.~Starobinsky,
Adv. Ser. Astrophys. Cosmol. \textbf{3}, 130-133 (1987)

\bibitem{Bezrukov:2008ut}
F.~Bezrukov, D.~Gorbunov and M.~Shaposhnikov,
JCAP \textbf{06}, 029 (2009)
doi:10.1088/1475-7516/2009/06/029
[arXiv:0812.3622 [hep-ph]].

\bibitem{Birrell:1982ix}
N.~D.~Birrell and P.~C.~W.~Davies,
``Quantum Fields in Curved Space,''

\bibitem{Enqvist:2015sua}
K.~Enqvist, S.~Nurmi, S.~Rusak and D.~Weir,
JCAP \textbf{02}, 057 (2016)
[arXiv:1506.06895 [astro-ph.CO]].

\bibitem{Repond:2016sol}
J.~Repond and J.~Rubio,
JCAP \textbf{07}, 043 (2016)
[arXiv:1604.08238 [astro-ph.CO]].

\bibitem{Ellis:2020krl}
J.~Ellis, D.~V.~Nanopoulos, K.~A.~Olive and S.~Verner,
[arXiv:2008.09099 [hep-ph]].

\bibitem{Sahni:1990tx}
V.~Sahni,
Phys. Rev. D \textbf{42}, 453-463 (1990)

\bibitem{Fields:2019pfx}
B.~D.~Fields, K.~A.~Olive, T.~H.~Yeh and C.~Young,
JCAP \textbf{03}, 010 (2020)
[arXiv:1912.01132 [astro-ph.CO]].

\bibitem{Maggiore:1999vm}
M.~Maggiore,
Phys. Rept. \textbf{331}, 283-367 (2000)
[arXiv:gr-qc/9909001 [gr-qc]].

\bibitem{Giovannini:1999bh}
M.~Giovannini,
Phys. Rev. D \textbf{60}, 123511 (1999)
[arXiv:astro-ph/9903004 [astro-ph]].

\bibitem{Giovannini:1999qj}
M.~Giovannini,
Class. Quant. Grav. \textbf{16}, 2905-2913 (1999)
[arXiv:hep-ph/9903263 [hep-ph]].

\end{thebibliography}
\end{document}